\documentclass[twocolumn,showpacs,preprintnumbers,aps,prl]{revtex4}
\usepackage{graphicx}
\usepackage{dcolumn}
\usepackage{bm}

\begin{document}
\preprint{}

\title{Storage and retrieval of light pulses in atomic media with ``slow'' and ``fast'' light}
\author{A. Lezama$^{1,2}$, A.M. Akulshin$^1$, A.I. Sidorov$^1$ and
P. Hannaford$^1$}

\affiliation{$^{1}$ARC Centre of Excellence for Quantum Atom Optics
and Centre for Atom Optics and Ultrafast Spectroscopy, Swinburne
University of Technology,
Hawthorn, 3122, Australia\\
$^{2}$Instituto de F\'isica, Facultad de Ingenier\'ia, C. Postal 30,
11000 Montevideo, Uruguay}%

\date{\today}

\begin{abstract}
We present experimental evidence that light storage, \textit{i.e.}
the controlled release of a light pulse by an atomic sample
dependent on the past presence of a writing pulse, is not restricted
to small group velocity media but can also occur in a negative group
velocity medium. A simple physical picture applicable to both cases
and previous light storage experiments is discussed.

\end{abstract}

\pacs{42.50.Gy, 32.80.Bx, 32.80.Qk}

\maketitle

All-optical information processing requires the use of photons as
fast and reliable carriers of information. Photons are quantum
objects and the search for media where the quantum state of photons
can be preserved and processed is of great significance. Recently,
broad attention has been focussed on the possibility of ``light
storage'' (LS) which is the preservation of the information carried
by a light pulse for controllable later release. Such a possibility
was suggested in theory \cite{FLEISCHHAUER00} and subsequent
experimental results were presented in support of this suggestion
\cite{PHILLIPS01,LIU01,ZIBROV02,MAIR02}. In all of these
experiments, a weak light pulse was ``written'' into an atomic
medium driven by a stronger field and, after a dark interval,
retrieved from the medium by turning on the strong (drive) field.

All observations of LS were achieved under conditions of
electromagnetically induced transparency (EIT) since, according to
\cite{FLEISCHHAUER00}, EIT and ``slow light'' (small group velocity
associated with EIT) play a key role. The storage effect is seen as
a consequence of the slowing and compression of the light pulse in
the atomic medium, the propagation of a mixed light-matter
excitation (dark-state polariton), the transformation of the
dark-state polariton in the absence of light into a pure atomic spin
excitation and finally the release of a light pulse once the drive
field is turned on \cite{LUKIN03}.

The purpose of this letter is to present experimental results and
theoretical considerations which broaden the scope of the subject by
demonstrating that a LS effect, analogous to that previously
reported \cite{PHILLIPS01,LIU01,ZIBROV02,MAIR02}, can take place in
media where EIT does not occur and where the probe pulse group
velocity is negative (``fast light'') as a result of large anomalous
dispersion.

Steep anomalous dispersion exists in driven atomic media in
connection with electromagnetically induced absorption (EIA)
\cite{AKULSHIN99}. EIA occurs when resonant light interacts with a
two-level atomic transition in which the Zeeman degeneracy of the
excited level is higher than that of the lower level; namely
$F_e>F_g>0$ where $F_e$ and $F_g$ are the total angular momenta of
the excited level and ground level, respectively. A resonant
increase in the probe absorption occurs under the condition of a
Raman resonance with ground state Zeeman sublevels
\cite{AKULSHIN98}. In particular, at zero magnetic field the EIA
resonance condition is achieved for the two orthogonal polarization
components of a single monochromatic optical field. Superluminar
pulse propagation in an atomic vapor under the conditions of EIA has
been demonstrated in \cite{AKULSHIN03}.

The experimental scheme used is very similar to the one presented in
Ref. \cite{PHILLIPS01}. LS was studied in a 5 cm long vapor cell
containing a natural isotopic mixture of rubidium. The cell was
placed at the center of a cylindrical $\mu $-metal shield. A
solenoid inside the magnetic shield allows tuning of the
longitudinal magnetic field $B$. The cell was heated ($\sim70\
{{}^\circ}$C) to produce almost $100\%$ linear absorption and
$50-80\%$ absorption at maximum light intensity. Two extended cavity
diode lasers were used for the $D_{1}$ and $D_{2}$ transitions. Fast
switching on and off of the laser light was achieved with an
acousto-optic modulator (AOM). After the AOM, a polarizer fixes the
polarization of the drive field. A Pockels cell along the light beam
was used as an electro-optic modulator (EOM) to generate a probe
pulse with orthogonal polarization relative to the drive field. The
probe pulse was 5
times weaker than the drive field. Care was taken to use the EOM in the $%
\lambda /2$ configuration in order to produce in-phase probe pulses
relative to the drive field. The light beam was expanded to a $1$ cm
diameter before the cell. The maximum laser power available at the
cell was $0.6$ mW and $2$ mW for the $D_{1}$ and $D_{2}$ lines,
respectively. After traversing the atomic vapor, the drive and probe
polarization components of the light were separated by a polarizing
beam splitter and collected by fast photodiodes. The electronic
control and detection response times were shorter than 1 $\mu s$.

\begin{figure}
\includegraphics[width=3.5in]{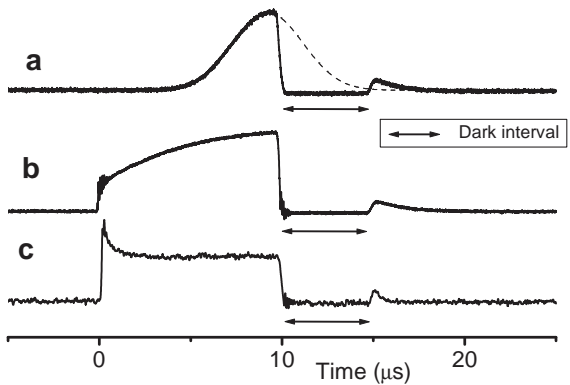}
\caption{\label{onlyexp}Observed signals for the probe field intensity transmission. a) Transition: $%
5S_{1/2}\left( F=2\right) \rightarrow 5P_{1/2}\left( F^{\prime
}=1\right) $ of $^{87}$Rb with Gaussian probe pulse; dashed:
Gaussian envelope of transmitted probe pulse without dark interval.
b) Same transition; square probe pulse. c) Transition:
$5S_{1/2}\left( F=2\right) \rightarrow 5P_{3/2}\left( F^{\prime
}=3\right) $ of $^{87}$Rb; square probe pulse.}
\end{figure}

Our setup allows the use of linear perpendicular or opposite
circular drive and probe field polarization combinations.
Qualitatively similar LS effects were observed for both choices. We
describe in the following the signals obtained with perpendicular
linear drive and probe fields polarizations.
LS under conditions of EIT was observed on both the transitions $%
5S_{1/2}\left( F=1\right) \rightarrow 5P_{3/2}$ ($D_{2}$ line) and
$5S_{1/2}\left( F=2\right) \rightarrow 5P_{1/2}\left( F^{\prime
}=1\right) $ ($D_{1}$ line) of $^{87}$Rb. We present results
obtained for the latter transition which was also used in previous
experiments \cite{PHILLIPS01,ZIBROV02,MAIR02}. The width of the EIT
resonance, measured at maximum light power by varying the magnetic
field, was 65 kHz. Trace \textbf{a} of Fig. \ref{onlyexp}, obtained
with a Gaussian shape probe pulse, reproduces the essential features
of the previous LS experiments
\cite{PHILLIPS01,LIU01,ZIBROV02,MAIR02}. Since the inverse of the
pulse duration is small compared to the EIT resonance width, the
Gaussian pulse shape is well preserved during propagation. A pulse
delay (relative to vacuum propagation) of approximately $1$ $\mu $s
is observed corresponding to a group velocity $v_g\sim 5\times 10^4$
m/s. We subsequently used square probe pulses to simplify further
comparison with numerical modelling. Trace \textbf{b} was obtained
for the same transition as for \textbf{a} with a square probe pulse.
The transmitted probe pulse presents a characteristic distortion,
indicative of a slow light medium \cite{AKULSHIN03}. In both traces
the retrieved pulse has an exponential decay with approximately the
same decay time ($\sim 2\ \mu $s). An exponential decay time of
$8\pm 2\ \mu $s was measured for the retrieved pulse amplitude as a
function of the dark interval, in agreement with the transverse
atomic time-of-flight estimate.

We now turn our attention to the $D_{2}$ line where EIA leads to
fast light propagation. Trace \textbf{c} of Fig. \ref {onlyexp} was
obtained with the laser tuned to the $5S_{1/2}\left( F=2\right)
\rightarrow 5P_{3/2}\left( F^{\prime }=3\right) $ transition of the
$D_{2}$ line. We notice a characteristic probe pulse distortion with
a leading edge less absorbed than the rest of the pulse. Such
distortion is indicative of the pulse advancement expected in a fast
light medium \cite {AKULSHIN03}. We observe that in the fast light
medium the retrieved pulse has similar characteristics to the pulse
retrieved from the slow light medium. In either case, in the absence
of a magnetic field, the retrieved pulse has an exponentially
decaying slope with a time constant dependent on the drive field
intensity. In the presence of a nonzero magnetic field similar
damped oscillations are seen on the retrieved pulse envelope for the
two types of transitions. The same dependence of the retrieved pulse
amplitude on the dark interval duration is observed in the two
cases.

\begin{figure}
\includegraphics[width=3.5in]{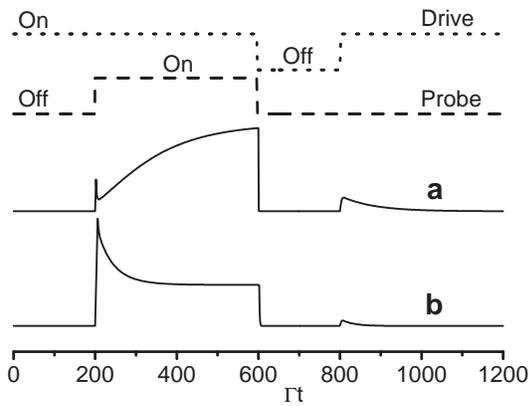}
\caption{\label{onlytheo}Dotted (dashed): drive (probe) time
sequence. a) Calculated transmitted probe field intensity for the
transition $5S_{1/2}\left( F=2\right) \rightarrow 5P_{1/2}\left(
F^{\prime }=1\right) $ of $^{87}$Rb. b) Same for the transition
$5S_{1/2}\left( F=2\right) \rightarrow 5P_{3/2}\left( F^{\prime
}=3\right)$. Parameters (see \cite{VALENTE02}):$\;\Omega _D/\Gamma
=0.3$ (a), $\Omega _D/\Gamma =0.6$
(b), $\Omega _D/\Omega _P=2$, $\gamma /\Gamma =10^{-3}$, $\alpha=8$, $B=0$ ($%
\Omega _D$ and $\Omega _P$ are the drive and probe field Rabi
frequencies).}
\end{figure}

To model the observed results, we have solved the Bloch equations
for an homogeneous ensemble of atoms with two energy levels with
Zeeman degeneracy \cite{VALENTE02}. The excited state decays
spontaneously to the ground state at a rate $\Gamma $ and, in order
to mimic time-of-flight relaxation, all states in the system decay
at rate $\gamma $ while ``fresh atoms'' are isotropically injected
in the ground levels at the same rate. The atoms interact with an
optical field whose polarization is decomposed into two orthogonal
(linear or circular) components with arbitrary amplitude and phase.
The light propagation in the atomic sample is considered to the
lowest order in the sample optical length \cite{WALSER94}.
Consequently, it is assumed that all atoms in the sample see the
same
(undepleted) incident field. Under such an assumption, the transmitted field is $%
\vec{E}_T\simeq \vec{E}_0+i\alpha\vec{P}$ where $\vec{E}_0$ is the
total incident field, $\vec{P}=Tr\left( \rho \vec{D}\right) $is the
atomic optical polarization per atom, $\vec{D}$ is the electric
dipole operator and $\alpha$ is a real constant proportional to the
atomic density and the sample length that is adjusted to the
observed absorption. The transmitted signal can be computed for
either polarization component. The pulse sequence used for the
calculation is shown in the upper traces of Fig. \ref{onlytheo}.
Trace \textbf{a} corresponds to the calculated transmission (at the
probe field polarization) for parameters corresponding to the
transition $5S_{1/2}\left( F=2\right) \rightarrow 5P_{1/2}\left(
F^{\prime }=1\right) $ of $^{87}$Rb. Trace \textbf{b} was calculated
for the transition $5S_{1/2}\left( F=2\right) \rightarrow
5P_{3/2}\left( F^{\prime }=3\right) $ of $^{87}$Rb. The pulse
distortions observed in the experiment are well reproduced and a
retrieved pulse is obtained for both transitions after the dark
interval. The amplitude of the retrieved pulse depends on the probe
pulse amplitude and duration and decays exponentially with a decay
time depending on the drive field intensity alone. Similar results
to those presented in Fig. \ref{onlytheo} are obtained if orthogonal
circular polarization are considered for both fields.

\begin{figure}
\includegraphics[width=3.5in]{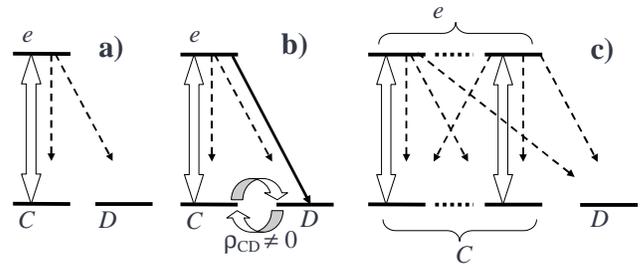}
\caption{\label{levelschemes}Basic level-schemes for the light atom
interaction. White arrow: applied field; dashed arrow: spontaneous
emission; solid arrow: stimulated Raman emission; curved arrow:
ground state coherence. a) Incoherent optical pumping, b) optical
pumping with coherence, c) generalization of a) for two levels with
Zeeman degeneracy.}
\end{figure}

The above results can be explained on the basis of a unique physical
picture. Consider the basic light-atom interaction process presented
in Fig.
\ref{levelschemes}a where an optical field couples the atomic ground-state $%
C $ (coupled state) to the excited state $e$ while the second
ground-state $D$ (dark state) is unaffected by the field. After
excitation in state $e$ the atom can decay spontaneously to both
lower states. This is an optical pumping process \cite {HAPPER72}.
If the atoms interact with the light for a sufficiently long time
and if states $C$ and $D$ are stationary, the atoms will eventually
end in state $D$ and the medium will become transparent. During this
process, transitions from state $e$ to state $D$ occur spontaneously
but may also take place through stimulated Raman emission. The
condition for stimulated Raman emission is the existence of nonzero
coherence between states $C$ and $D$, \textit{i.e.} $\rho _{DC}\neq
0$ where $\rho _{DC}$ is the element of the density matrix between
states $D$ and $C$. Under such a condition the interaction with the
applied field results in optical coherence between states $e$ and
$D$ ($\rho _{eD}\neq 0$) which in turn radiates a field for this
transition which is coherent with the initial field. As a general
rule, \emph{a light field interacting with a coherently prepared
atom in the ground state will induce coherent emission of light in
all allowed optical transitions as long as the (ground state) atomic
coherence is preserved} (Fig. \ref{levelschemes}\textbf{b}).

Direct generalization of the situation represented in Fig \ref{levelschemes}%
\textbf{a} to the more complex case of a two-level system with
arbitrary Zeeman degeneracy interacting with an optical field is
possible. In general, given the field polarization, after a suitable
basis transformation, one can identify several ($C$) states in the
ground level, each coupled to a corresponding excited state
sublevel, and two, one or zero dark ($D$) states (Fig.
\ref{levelschemes}\textbf{c}). The last case occurs when the Zeeman
degeneracy of the excited level is higher than that of the ground
level \cite{SMIRNOV89,TAICHENACHEV04}. In all cases the general rule
stated above applies.

Two types of systems have commonly been considered for the
experimental study of coherent light-atom interaction dynamics. In
the first class one has the Hanle type experiments \cite
{KASTLER73,HANNAFORD97,DANCHEVA00,VERKERK01,FAILACHE03} where a
single optical field with well defined polarization interacts with
two atomic levels with Zeeman sublevels whose energy is tuned with a
static magnetic field. If the light polarization is linear it is
straightforward that a scheme similar to that of Fig.
\ref{levelschemes}\textbf{c} results by choosing the quantization
axis along the direction of the light polarization. Even in the
general case of arbitrary elliptical light polarization the scheme
of Fig. \ref{levelschemes}\textbf{c} can be applied after a suitable
transformation of the state basis that depends on the incident light
polarization \cite {TAICHENACHEV04}. An immediate consequence of
this description is that in the presence of ground state coherence
the system will respond by the coherent emission of light with a
polarization orthogonal to that of the incident field. Such a
process was recently discussed in terms of the polarization change
that the light experiences while propagating in an atomic medium
with light induced anisotropy and an alternative explanation of the
LS effect in such terms was proposed \cite{KOZLOV04}.

A second class of experiments for which the picture presented in
Fig. \ref{levelschemes}\textbf{a}-\textbf{b} can be conveniently
applied, concerns a system of an excited level and two ground levels
with energy separation $\Delta $ driven by two optical fields of
frequency $\omega _1$ and $\omega _2$. This system (hereafter
designated as the $\Lambda $ system), has been extensively
studied theoretically and experimentally in connection with EIT. Since in a $%
\Lambda $ system each field interacts with a different transition,
it is possible after a time dependent unitary transformation to
describe the atomic dynamics by a time-independent Hamiltonian and
to identify a dark state $D$ and a coupled state $C$
\cite{SCULLYBOOK97}. This ``coupled-uncoupled'' description
 exactly corresponds to the situation of Fig.
\ref{levelschemes}\textbf{a}-\textbf{b}. Both levels $C$ and $D$ are
linear combinations of the two lower levels that explicitly depend
on the amplitudes and phases of the
applied fields. Here again, if coherence is present between levels $C$ and $%
D $, the atomic medium will react to the optical excitation by the
coherent emission of fields at frequencies $\omega _1$ and $\omega
_2$. These fields
are ``orthogonal'' to the applied fields in the sense that they \emph{%
together} only couple to state $D$ (while the incident fields only
couple to $C$). As an example we consider the situation
corresponding to the retrieval process in the LS experiment reported
in \cite{LIU01}. After the dark interval, the drive field ($\omega
_1$) alone excites the $\Lambda $ system. Since (previously created)
coherence is present between the ground levels
(that coincide with $C$ and $D$ in this case), a field of frequency $%
\omega _2=\omega _1-\Delta $ is emitted by the medium. In the
general case, in the presence of coherence between $C$ and $D$, two
fields will be emitted with frequencies $\omega _1$ and $\omega _2$
that will interfere with the incident fields.

The previous discussion leads to a very general qualitative
understanding of the transient behavior of coherently driven atomic
systems. For given excitation conditions, after a long enough time,
the system will reach a steady state. If one or more dark states
exists the steady state of the system will be the dark state(s) $D$.
If no dark state exists the steady state will generally be a
statistical mixture (diagonal density matrix) of states $C$. A rapid
modification of the excitation conditions, for instance a light
polarization change in a Hanle experiment or a change in the
amplitude
or phase of the fields in a $\Lambda $ scheme, will determine a new set ($%
C^{\prime }$, $D^{\prime }$) of coupled and dark states. The change
in the excitation conditions needs to be rapid with respect to the
optical pumping time but may otherwise be slow with respect to other
characteristic times. Since the previous
state of the system is generally a linear combination of the new $C^{\prime }$, $%
D^{\prime }$ states, coherence among the latter states exists and
results in the transient emission of an ``orthogonal'' field. Such
emission will last as long as the coherence between $C^{\prime }$
and $D^{\prime }$ survives. The decay of the transient emission will
be purely exponential with a time
constant given by the optical pumping time between states $C^{\prime }$ and $%
D^{\prime }$ if those states are stationary. If they are not
stationary (nonzero
magnetic field for a Hanle experiment or nonzero Raman detuning in a $%
\Lambda $ system), the decay will show damped oscillations
\cite{VALENTE02}. If the two different excitation conditions are
separated in time by a dark interval then the transient emission of
an ``orthogonal'' field will occur after the dark interval provided
that this interval is not long compared to the ground state
coherence lifetime. This mechanism applies to the LS experiments
previously reported \cite {PHILLIPS01,LIU01,MAIR02,ZIBROV02}.

The experimental results and the discussion above demonstrate that
EIT and propagation in a ``slow light'' medium is not an essential
requirement for the storage and retrieval of a light pulse in an
atomic medium. One point in the simple description presented here,
which departs from previous theoretical treatments
\cite{FLEISCHHAUER00}, is the simple consideration of the field
propagation in the atomic medium where the spatial variation of the
incident field along the sample is essentially neglected. This crude
approximation is nevertheless appropriate under conditions where the
duration of the probe pulse is comparable or longer than the light
propagation time. Such a situation occurred in most LS experiments
reported so far (including ours) with the exception of \cite
{LIU01}. In the picture presented here, the light retrieval
transient appears as a consequence of the irreversible relaxation of
the system towards a new steady state. In this context, only
exponential decaying pulses can be obtained preventing the retrieval
of information about the state of the initial probe pulse other than
its presence (one bit information). The storage and recovery of
information on the state of the incoming probe pulse are beyond the
scope of this simple picture and possibly of most experimental
conditions achieved to date.

In conclusion, we have achieved storage and retrieval of light
pulses in slow-light and fast-light media and given a unified
description of both cases. The results presented in this letter
suggest that further theoretical and experimental work is needed for
the understanding and practical realization of information
preserving storage in an atomic medium. A necessary first step in
this direction would be the realization of an atomic medium in which
more than one probe pulse (more than one bit of information) can be
contained in the atomic sample and propagate without significant
distortion \cite{BOYD05,MATSKO05}.

This work was supported by a Swinburne University RDGS grant and
Fondo Clemente Estable (Uruguay).

\end{document}